\documentclass[superscriptaddress,pre, showkeys, bibtex, 11pt]{revtex4-1}

\usepackage[T1]{fontenc}
\usepackage[latin9]{inputenc}
\usepackage{hyperref}

\usepackage[left=2.5cm, right=1.5cm, top=2cm, bottom=2cm]{geometry}

\usepackage[table,svgnames]{xcolor}
\usepackage{array}
\usepackage{units}
\usepackage{amsmath}
\usepackage{amssymb}
\usepackage{graphicx}

\begin{document}

\title[Feedback mechanisms for self-organization]{Feedback mechanisms for self-organization to the edge of a phase transition} 

\author{Victor Buend{\'\i}a}
\affiliation{Departamento de Electromagnetismo y F{\'i}sica de la Materia and\\ Instituto Carlos I de F{\'i}sica Te{\'o}rica y Computacional. Universidad de Granada, E-18071 Granada (Spain)}
\affiliation{Dipartimento di Matematica, Fisica e Informatica, Universit\`a di Parma, via G.P. Usberti, 7/A-43124, Parma, Italy}
\affiliation{INFN, Gruppo Collegato di Parma, via G.P. Usberti, 7/A-43124, Parma, Italy}
\author{Serena di Santo}
\affiliation{Morton B. Zuckerman Mind Brain Behavior Institute Columbia University, NY, USA}
\author{Juan A. Bonachela }
\affiliation{Department of Ecology, Evolution, and Natural Resources, Rutgers University, New Brunswick, 08901 NJ (USA)}
\author{Miguel A. Mu{\~n}oz}
\email{mamunoz@onsager.ugr.es}
\affiliation{Departamento de Electromagnetismo y F{\'i}sica de la Materia and\\ Instituto Carlos I de F{\'i}sica Te{\'o}rica y Computacional. Universidad de Granada, E-18071 Granada (Spain)}
\affiliation{Dipartimento di Matematica, Fisica e Informatica, Universit\`a di Parma, via G.P. Usberti, 7/A-43124, Parma, Italy}

\begin{abstract}
Scale-free outbursts of activity are commonly observed in physical, geological, and biological systems.  The idea of self-organized criticality (SOC), introduced back in 1987 by Bak, Tang and Wiesenfeld suggests that, under certain circumstances, natural systems can seemingly self-tune to a critical state with its concomitant power-laws and scaling. Theoretical progress allowed for a rationalization of how SOC works by relating its critical properties to those of a standard non-equilibrium second-order phase transition that separates an active state in which dynamical activity reverberates indefinitely, from an absorbing or quiescent state where activity eventually ceases. The basic mechanism underlying SOC is the alternation of a slow driving process and fast dynamics with dissipation, which generates a feedback loop that tunes the system to the critical point of an  absorbing-active continuous phase transition. Here, we briefly review these ideas as well as a recent closely-related concept: self-organized bistability (SOB). In SOB, the very same type of feedback operates in a system characterized by a discontinuos phase transition, which has no critical point but instead presents bistability between active and quiescent states. SOB also leads to scale-invariant avalanches of activity but, in this case, with a different type of scaling and coexisting with anomalously large outbursts. Moreover, SOB explains experiments with real sandpiles more closely than SOC. We review similarities and differences between SOC and SOB by presenting and analyzing them under a common theoretical framework, covering recent results as well as possible future developments.  We also discuss other related concepts for ``imperfect'' self-organization such as ``self-organized quasi-criticality'' and ``self-organized collective oscillations'', of relevance in e.g. neuroscience, with the aim of providing an overview of feedback mechanisms for self-organization to the edge of a phase transition.
\end{abstract}


\maketitle

\section{Introduction}
The seminal work of Bak, Tang, and Wiesenfeld in which the idea of ``self-organized criticality'' was first introduced \cite{BTW}, which has been cited thousands of times in the scientific literature and beyond, opened a whole research field and triggered a huge avalanche of scientific excitement in Statistical physics. Fractals \cite{Mandelbrot} can be considered as precursors of these ideas, and scale-free complex networks \cite{Barabasi} successors in the timeline of waves of scientific interest.

Bak and collaborators developed the groundbreaking idea that scaling behavior is observed in Nature owing to self-organization mechanisms that tune systems to the vicinity of critical points \cite{Bak}.  Thus, self-organized criticality (SOC) helped shed light onto why scale-invariant phenomena (both in space and time) are so commonly observed in natural systems, in spite of the fact that criticality, i.e. second-order phase transitions, with their associated power-laws and scaling, occur only at singular (critical) points of parameter spaces \cite{BTW,Bak,Bak-Chen} (for pedagogical reviews and detailed accounts of SOC, we refer to \cite{Jensen,Moloney,Pruessner,BJP,SOC-25,Turcotte,Gros}).

The most succesful and archetypical examples of SOC are sandpile toy models \cite{BTW,Manna,Oslo,Zhang,Grassberger-Manna,MZ}\footnote{Alternative models and mechanisms such as, for example, the celebrated Bak-Sneppen model for punctuated evolution relying on ``extremal dynamics'' \cite{Bak-Sneppen} were also proposed to achieve scaling in a self-organized way, but we will not discuss them here.}. In sandpiles, ``grains'' --which represent in an abstract way some token of ``stress'' or ``energy'' \cite{Early1}-- are slowly added into a system (usually a lattice or another type of network), and locally redistributed on a fast way whenever an instability threshold is overcome. This redistribution triggers avalanches of topplings, eventually dissipating some of these grains at the system's open boundaries. Upon iteration, these dynamics result in the self-organization of the system to a critical stationary state that exhibits power-law avalanche distributions and obeys finite-size scaling \cite{Bak,Jensen,Moloney,Pruessner,Dhar99,Dhar06,Turcotte,GG,BJP,Laurson}.

The observation of scale invariance and other features characteristic of criticality without the need for parameter fine tuning prompted an enormous interest in these simple models. As a word of caution, let us remark that it was also soon realized that sandpile models bear little resemblance with the physics of actual sandpiles as experimentally analyzed in the laboratory. In actual sandpiles, ingredients such as inertia, gravity, and stickiness (typically absent in standard SOC models) play important roles, and scale invariance is not easily observed \cite{Pruessner,Turcotte,BJP}. Empirical evidence of SOC is more easily found in ricepiles or in superconductors (see \cite{Pruessner} for an account on experimental realizations as well as for other general aspects of SOC). Let us just highlight that compelling evidence of SOC has been recently found in an ultracold atomic gas \cite{Ultracold,Ultra2}). This discovery illustrates that, more than $30$ years after its birth, SOC is still a powerful, relevant and pervading concept.

On the theoretical side, a key ingredient of the mechanism for self-organization in sandpiles is the fact that driving and dynamics operate at two broadly separated timescales (i.e. slow-fast dynamics) \cite{Bak,Jensen,Moloney,GG}. An infinite separation of timescales is usually achieved by driving the system only when all activity has stopped, but not during avalanches (``\textit{infinitely slow}'' or ``\textit{offline}'' \textit{driving}); if this is not the case, a finite characteristic (time/size) scale appears \cite{GG,GG2,GG3}.  Similarly, \textit{conservative dynamics} in the bulk of the system are also key to SOC, because bulk dissipation leads necessarily to the emergence of characteristic spatio-temporal scales, thus preventing the possibility of scale-invariance \cite{GG,Bak-non,Hwa1,Hwa2,Drossel-nonconserved,Malcai}. We refer to \cite{GG} for a more in depth theoretical discussion on the emergence of generic scale invariance, conservation laws, and SOC.

A large variety of sandpile models, with diverse microscopic rules, were investigated after the original proposal of Bak and colleagues (see a compilation of prototypical SOC models in \cite{JABO1} and \cite{Jensen,Moloney,Pruessner}). The main additional ingredient was the introduction of stochasticity in the redistribution rules, replacing the fully deterministic updating rules of the original sandpile \cite{Manna}. Given the diversity of models, a compelling question emerged as to whether there is universality in SOC (i.e. models/systems that share the same scaling features) \cite{Ben-Hur,Thesis-JA}. From the computational viewpoint it soon became clear that, in spite of preliminary evidence, the original (deterministic) sandpile model of Bak, Tang and Wiesenfeld (BTW) does not obey clean scaling behavior but rather some type of multiscaling or anomalous scaling \cite{anomalies1,anomalies2,anomalies3,anomalies4}. This anomaly stems from the breaking of ergodicity \cite{Grassberger-Manna}, and the existence of many conservation laws associated with the deterministic nature of its updating rules\footnote{Let us remark that there exist very powerful theoretical tools for deterministic (Abelian) sandpiles \cite{Dhar99,Dhar06}, a theoretical endeavor complementary to the type of analyses for stochastic sandpiles discussed here.}. On the other hand, sandpiles with some level of stochasticity (such as the Manna model \cite{Manna} or the Oslo ricepile model \cite{Oslo}) exhibit standard and universal scaling behavior, even though large-scale simulations and careful computational analyses were required to reach such a conclusion (see e.g. \cite{Christensen-2004,Pruessner-universality,JABO1,Grassberger-2016,JABO-MZ}).

 Because criticality and universality are hallmarks of second-order phase transitions, diverse attempts were made to map the behavior emerging in SOC systems to that of standard (continuous or second-order) phase transitions. In particular:
\begin{itemize}
\item A first proposal mapped sandpiles to the pinning-depinning transition of interfaces moving in random media \cite{Interfaces1,Interfaces2,Interfaces3,Interfaces4,Pruessner-Oslo}.  In this approach, the height of the interface at a given location corresponds to the number of times that such a site has toppled in the sandpile. This successful mapping has profound physical implications, as pinning-depinning transitions are also related to the dynamics of magnetic domain walls in random media, the Barkhausen effect, and $1/f$ noise, which had long been studied and are known to display scale invariance \cite{Sethna1,Sethna2,Barkhausen}.

\item A second proposal, on which we focus here, connected SOC with reaction-diffusion systems exhibiting absorbing-active phase transitions \cite{BJP,FES0,FES-PRE,FES-PRL,Romu-PRE,Romu-PRL,JABO1}. The mapping was proposed on general symmetry and conservation principles, and afterward refined in an exact formal way \cite{Wiese-Manna}.  \end{itemize}

These two apparently disparate approaches were found to be fully equivalent to each other, first using heuristic and numeric arguments \cite{Alava-Munoz,JABO-PRL,JABO-cusps} and then with more rigorous analyses \cite{Wiese-mapping}.

In order to scrutinize how SOC behavior is related to standard phase transitions, the notion of ``fixed-energy sandpiles'' (FESs) was introduced, an idea similar in spirit to an early suggestion by Tang and Bak \cite{Early1,Early2}.  The key idea was to ``regularize'' sandpiles by switching off both slow driving and boundary dissipation, with the total number of sandgrains in the system thus becoming fixed, i.e. a conserved quantity, suitable to be considered a control parameter \cite{BJP,FES0,FES-PRE,FES-PRL,FES1}. Thus, the state of a FES is described by two quantities: the total number of sandgrains in the system (control parameter) and the total number of sites that are above the threshold of instability (order parameter). The latter is based on the fact that, in a sandpile, sandgrains can be either ``active'' if they happen to be above threshold (ready to topple and be redistributed), or ``inactive'' otherwise. Inactive grains can, however, contribute to future activations. Using a more general and abstract language, ``{\bf energy}'' hereon refers to the mean accumulated stress (e.g. total number of sandgrains per site on the sandpile) while ``{\bf activity}'' describes the number of sandgrains above the instability threshold.

Not surprisingly, FESs exhibit two distinct phases depending on the value of their energy $E$: either they are in an ``active'' phase with ceaseless redistribution of activity for sufficiently large values of $E$, or they are in an absorbing or quiescent phase in which all activity ceases and the dynamics are frozen \cite{Haye,Marro,Henkel} (see Figure 1, left panel). Thus, there exists a \textit{continuous} absorbing-to-active phase transition at a critical energy value $E_c$. Let us note that the existence of such a phase transition in FESs has only recently been demonstrated mathematically \cite{Sido,Ron-ARW}.

This observation allowed for a rationalization of SOC as a dynamical feedback mechanism that tunes the system to the edge of an absorbing-to-active phase transition through (slow) driving and (fast) bulk dynamics, occurring at infinitely separated timescales with boundary dissipation \cite{VZ,FES-PRE,FES-PRL,BJP,JABO1,Ivan,Broker2}. In other words, the steady state reached spontaneously by the SOC dynamics is characterized by an average steady-state energy $E^{SOC}$ such that $E^{SOC} =E_c$. As a consequence, the scaling features of SOC systems can be inferred from those of their corresponding fixed-energy counterparts using the powerful set of theoretical tools available for standard non-equilibrium phase transition.

Non-equilibrium phase transitions into absorbing states have long been studied, and it is well stablished that most of them share the same type of universal behavior, belonging to the so-called ``directed percolation'' (DP) universality class \cite{Marro,Haye,Henkel,Odor}. As in some DP systems, in FESs there is not one but many absorbing states. Any configuration with vanishing activity and arbitrary values of the energy is absorbing \cite{ManyAS}. However, in FESs there is an additional conservation law that might be relevant for universality issues (see below).

To help clarify this and other issues, here we use the formalism of Langevin equations to review classic and state-of-the-art theoretical aspects of SOC. This formalism follows the philosophy of the extremely succesful approach of Landau and Ginzburg to equilibrium phase transitions and critical points \cite{Binney,LeBellac,Amit}, as well as its extension to dynamical problems (as reviewed by Hohenberg and Halperin \cite{HH}). For each case, we will present the simplest (Langevin) equation, including the main symmetries, conservation laws, and stochastic effects present in the system, and neglecting irrelevant terms \cite{Binney,LeBellac,Amit}. This approach places the focus on universal scaling features, leaving aside unimportant microscopic details. Thus, such Langevin equations constitute an ideal starting point for further theoretical analyses (such as renormalization group calculations and other field theoretical approaches) and even for numerical studies. After presenting and discussing the theory of SOC, we move on to discussing related theories of self-organization to the edge of a phase transition. We next describe the theory for the self-organization to the edge of a discontinuous phase transition with bistability, and finally we discuss theories for ``imperfect self-organization'' either to a continuous or to a discontinuous transition. The latter can be of more relevance than the original self-organization theories to describe real-world situations.

\section{Theory of self-organized criticality (SOC)}
Let us start by  discussing the simplest possible SOC system \cite{SOBP}. For a macroscopic (mean-field) description of a sandpile, two relevant variables are needed: the overall energy $E$ (which represents the total density of sandgrains in sandpiles and is conserved in the bulk), and the overall activity $\rho$ (i.e. the density of sites which are above threshold). To analyze the possible connection between sandpiles and standard non-equilibrium phase transitions at a mean-field level, let us consider the simplest possible equation describing a continuous absorbing-active phase transition for the overall density $\rho$:
\begin{equation}\dot{\rho}(t) = a \rho(t) - b \rho^2(t)
\label{DP-MF}
\end{equation} 
where $a$ and $b>0$ are constants, and the fine-tuning of $a$ controls the behavior of the system.  This equation exhibits an absorbing phase with vanishing activity ($\rho=0$) below the critical point, i.e. for $a< a_c =0$, and an active phase with steady-state density $\rho=a/b\neq 0$ for $a>a_c=0$.

To establish the connection with SOC, let us start by linking the equation above with FESs. To that end, it is required an additional conserved energy (or energy density) $E$ such that it fosters the creation of activity (i.e. increases $a$ in Eq.(\ref{DP-MF})). Thus, in first approximation we can write:

\begin{equation} \dot{\rho}(t) = (a + \omega E) \rho(t) - b \rho^2(t) \label{FES-MF} \end{equation} 

where $\omega>0$ is simply a proportionality constant. Observe that, since $E$ is a conserved quantity (i.e. $\dot{E}=0$), it can be used as a control parameter keeping $a$ fixed.  In particular, the critical point lies now at $E_c=-a/\omega$. Equation (\ref{FES-MF}) constitutes the mean-field description of fixed energy sandpiles: a dynamical equation for the overall activity, $\rho(t)$, whose steady state is determined by the control parameter, the energy density $E$ in the system.

On the other hand, in the SOC version of sandpiles $E$ becomes a dynamical variable $E(t)$, which  increases by external driving (at an arbitrarily small rate $h$) and decreases owing to activity-dependent dissipation (at a rate $\epsilon \rho$). This can be summarized by the equation:
\begin{equation}
  \dot{E}(t)= h -\epsilon \rho(t).
  \label{feedback}
\end{equation}
In the double limit $h,\epsilon \rightarrow 0^+$ (infinitely separated timescales), or if $h/\epsilon \rightarrow 0$ (energy conservation), the steady-state solution of the system represented by Eqs.(\ref{FES-MF}) and (\ref{feedback}) is $\rho= h/\epsilon \rightarrow 0^+$ and $E^{SOC}= (b h/\epsilon -a)/\omega \rightarrow E_c$ (see Figure 1, left). In other words, the system self-organizes to the critical point of a standard absorbing-active phase transition, i.e. the critical state is a dynamical attractor of the system \cite{Early1}.

\begin{figure}[h] \begin{center} \includegraphics[height=5cm]{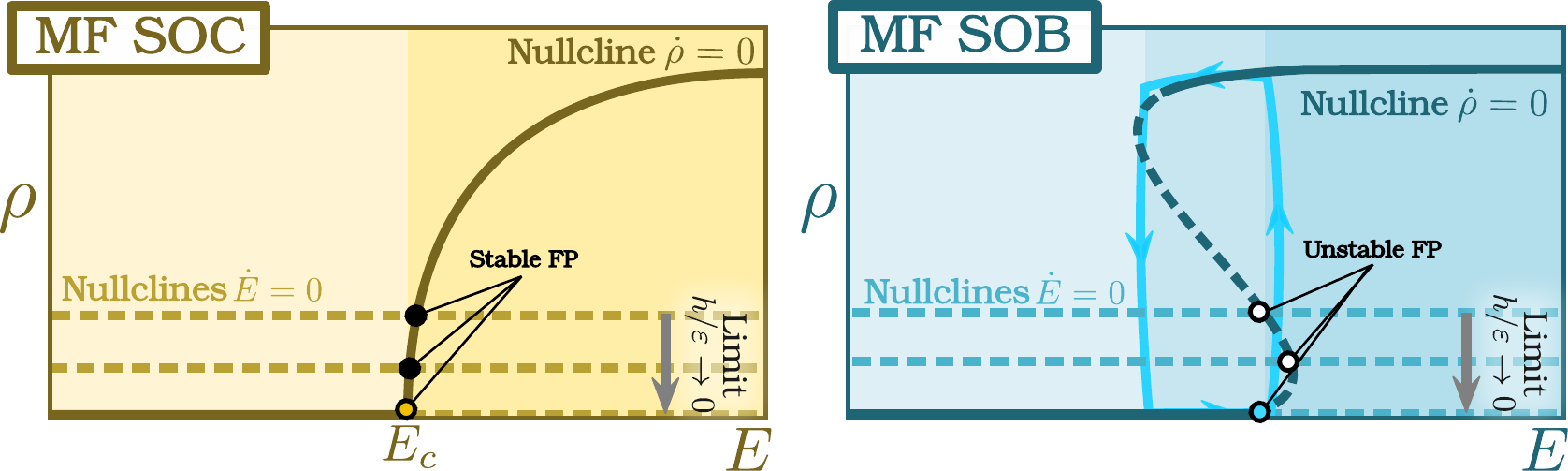} 

\caption{\textbf{Mean-field (MF) pictures of SOC and SOB.}  Sketch of the nullclines associated with the two dynamical mean-field equations defining self-organized criticality (SOC, panel A) and self-organized bistability (SOB, panel B). In both cases, the nullcline for the second (feedback) equation is plotted for three different values of the ratio $h/\epsilon$. (A) In the case of SOC, nullclines intersect at a stable fixed point, which becomes closer and closer to the critical point as the limit of infinite separation of timescales $h/\epsilon \rightarrow 0$ is approached (see grey arrow). (B) In the case of SOB, for sufficiently low values of $h/\epsilon$ the intersection between nullclines occurs on the so-called spinodal line (dashed dark line). Points located in the spinodal line are unstable, and the system presents a stable, fixed-amplitude limit cycle sketched by the cyan trajectory.} \label{nullclines} \end{center}       
\end{figure}
 
Observe that the key to the SOC mechanism lies on the feedback created by the dynamics of the control parameter $E$ (see Fig.\ref{nullclines}).  Its dynamics strongly depend on the system state/phase: if activity vanishes, $\rho=0$ and $\dot{E}= h$, leading to an increase in $E$ that shifts the system towards its supercritical phase. If, on the other hand, $\rho>0$, since $\epsilon \gg h$, then $\dot{E} \approx -\epsilon \rho$ and $E$ decreases, pushing the system towards the subcritical phase.  This feedback loop necessarily drives the system to the vicinity of the critical point, and exactly to the critical point if the separation of timescales is infinite, as shown above. In more general terms: \textit{the existence of a control mechanism that acts differentially on each phase --i.e. at each side of the phase transition-- creates a feedback loop that self-organizes the system to the very edge of the transition} \cite{GG2,BJP,JABO1} (see \cite{Sontag} for a discussion of this general idea in the context of control theory).

In order to go beyond this simple mean-field description, we need to extend the theory to make it spatially explicit and stochastic, i.e. shift from mean-field theory to stochastic field theory \cite{Amit,LeBellac}. The simplest possible equation describing absorbing phase transitions is the so-called Reggeon field theory (or DP theory), which can be written as the following Langevin equation \cite{RFT1,RFT2,Henkel}: 
\begin{equation}
  \partial_t \rho(\vec{x},t) = a \rho({\vec{x}},t)-b\rho^{2}({\vec{x}},t)
  +D\nabla^2\rho ({\vec{x}},t)+\sigma  \sqrt{\rho(\vec{x},t)}\eta({\vec{x}},t) 
\label{RFT}
\end{equation}
where $\rho({\vec{x}},t)$ is the activity field, $a$ and $b>0$ are constants, and $D$ and $\sigma$ are the diffusion and noise constants, respectively. $\eta (\vec{x},t)$ is a zero-mean Gaussian noise with $\langle \eta({\vec{x}},t) \eta({\vec{x'}},t) \rangle =\delta(\vec{x}-\vec{x'}) \delta(t-t') $ which, together with the prefactor $ \sqrt{\rho(\vec{x},t)}$, accounts for demographic fluctuations in particle numbers. Importantly, the noise term vanishes in the  absorbing state $\rho(\vec x, t) = 0$.
 
In analogy with the mean-field approach, we now use the equation above to represent FESs, for which we need to add another equation for the (conserved) energy coupled linearly with the activity \cite{FES-PRE,FES-PRL,BJP}:
 \begin{equation} \begin{array}{lll}
  \partial_t \rho(\vec{x},t) = (a + \omega E({\vec{x}},t))    \rho-b\rho^{2}
  +D\nabla^2\rho+\sigma  \sqrt{\rho(\vec{x},t)}\eta({\vec{x}},t) \\ 
\partial_{t} E({\vec{x}},t) = \nabla^2\rho({\vec{x}},t) \label{FES}
\end{array}
\end{equation}
where $E({\vec{x}},t)$ is the energy field. Some dependencies on $({\vec{x}},t)$ have been omitted for the sake of simplicity.  Note that the equation for the energy is diffusive, describing the redistribution of energy among neighboring locations with no loss in the presence of activity. Thus, the system-averaged energy per site (i.e. the spatial integral of the energy field divided by the system volume) $\overline{E}$ is constant in FESs and can be taken as a control parameter.  As in the case of the mean-field theory, Eqs.(\ref{FES}) exhibit a phase transition at a particular value of the average energy density: for $\overline{E}>E_{c}$ there are continuous ongoing redistributions of activity and energy, while for $\overline{E}<E_c$ the system eventually falls into the absorbing state $\rho({\vec{x}},t)=0$ (see e.g. \cite{Dornic}). The set of equations for FESs, Eqs.(\ref{FES}), was proposed on phenomenological grounds \cite{FES-PRE,FES-PRL} (see also \cite{PMB}) and later derived from a discrete reaction-diffusion model with many absorbing states and a local conservation law \cite{Romu-PRL}.  Only recently has it been derived in a rigorous way from the microscopic rules of a stochastic (fixed-energy) sandpile \cite{Wiese-Manna}.

Eqs.(\ref{FES}) can be integrated ``a la SOC'', e.g. by adding at the initial time and after each avalanche a discrete amount of energy and activity (``infinitely slow'' or ``offline'' driving), and considering open boundary conditions (i.e. allowing for boundary dissipation). The resulting self-organized system converges to the critical point of Eqs.(\ref{FES}). Alternatively, a continuous version can be achieved by including in Eqs.(\ref{FES}) an explicit (``online'') driving and a dissipation term:

\begin{equation} \begin{array}{lll}
  \partial_t \rho(\vec{x},t) = (a + \omega E({\vec{x}},t))    \rho-b\rho^{2}
  +D\nabla^2\rho+\sigma  \sqrt{\rho(\vec{x},t)}\eta({\vec{x}},t), \\ 
                   \partial_{t} E({\vec{x}},t) = \nabla^2\rho({\vec{x}},t)
-\epsilon \rho({\vec{x}},t) + h(\vec{x},t).
                   \label{SOC}
\end{array}
\end{equation}

Note that the small driving $h(\vec x, t)$ could also be added to the activity in order to avoid the absorbing state (in which the dynamics stop). Otherwise, a small seed of activity needs to be added to slightly perturb the system every time the absorbing state is reached. 
Note that this ``online'' methods are slightly different from the ``offline'' driving since the average energy field changes during avalanches and not only between them. 

\begin{figure}
\begin{center}
\includegraphics[height=8cm]{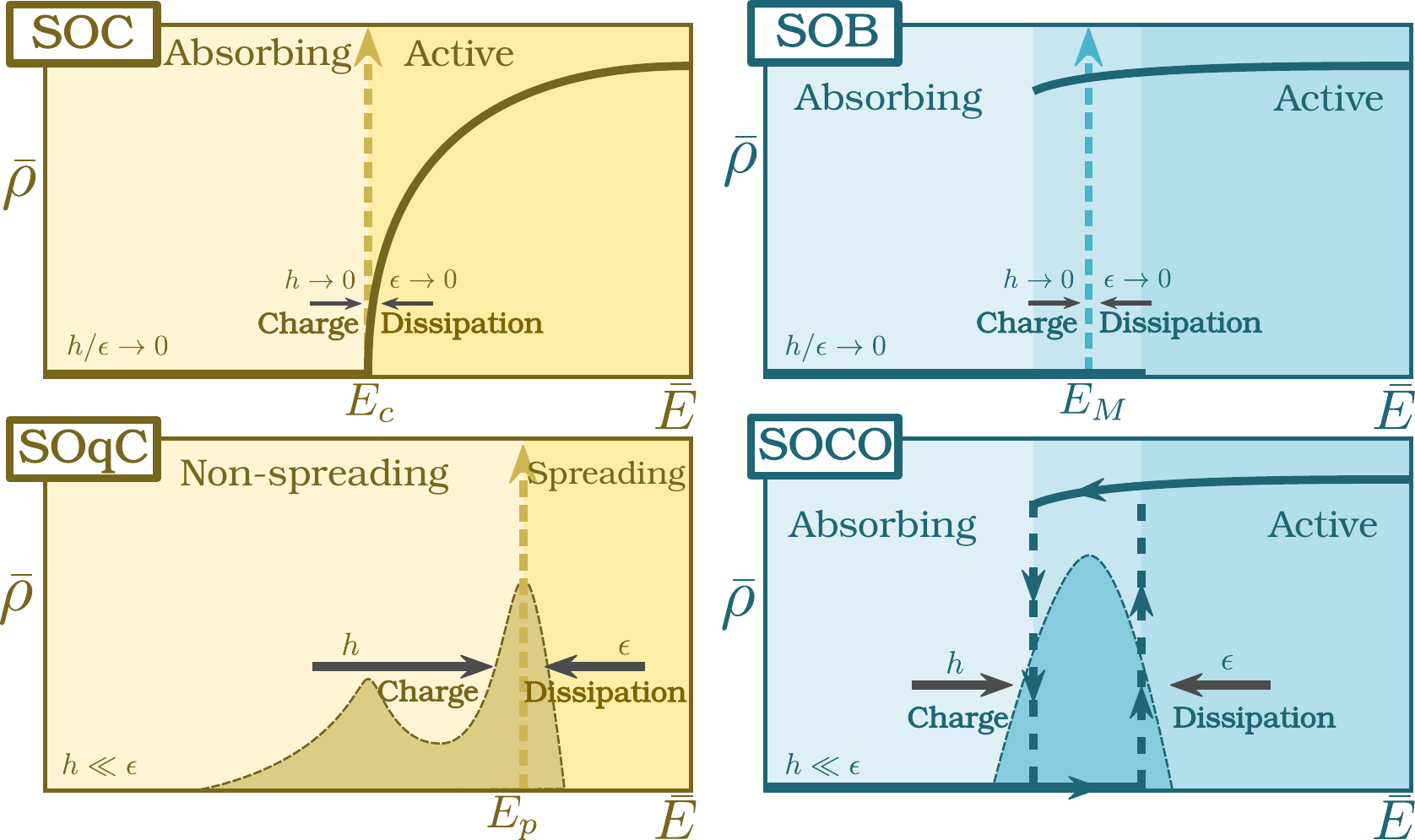}
\caption{\textbf{Sketch of the different types of self-organization mechanisms discussed here.}  The four panels illustrate, respectively, the mechanisms for self-organization to a continuous phase transition with criticality (left panels) and to a discontinuous transition (right panels), for both the ``perfect'' conserved case (top panels) and the ``approximated'' or ``imperfect'' non-conserved case (bottom panels).  In all cases the steady-state average activity is plotted as a function of the control parameter (the average ``energy'' or ``stress''). The SOC and SOB mechanisms change dynamically the control parameter to a precise value (either $E_c$ or $E_M$, respectively), meaning that the system becomes perfectly self-organized to the edge of a phase transition (either a continuous one for SOC or discontinuous one for SOB) in the thermodynamic limit.
    On the other hand, their corresponding ``imperfect'' or non-conserved counterparts --that we call ``self-organized quasi-criticality'' (SOqC) and ``self-organized collective oscillations'' (SOCO)-- give rise to broad distributions of possible energy values, even in the thermodynamic limit, typically around the edge of the transition point (shown as an area enclosed by dashed lines).  The thin arrows in the upper panels illustrate the fact that dissipation and driving rates are very small ($h\rightarrow 0$, $\epsilon \rightarrow 0$ with $h/\epsilon \rightarrow 0$), while the thick ones indicate that such a strict limit is not taken, but still $h \ll \epsilon$). }
\label{Panels}
\end{center}
\end{figure}

\begin{figure}[h]
\begin{center}
\includegraphics[height=8cm]{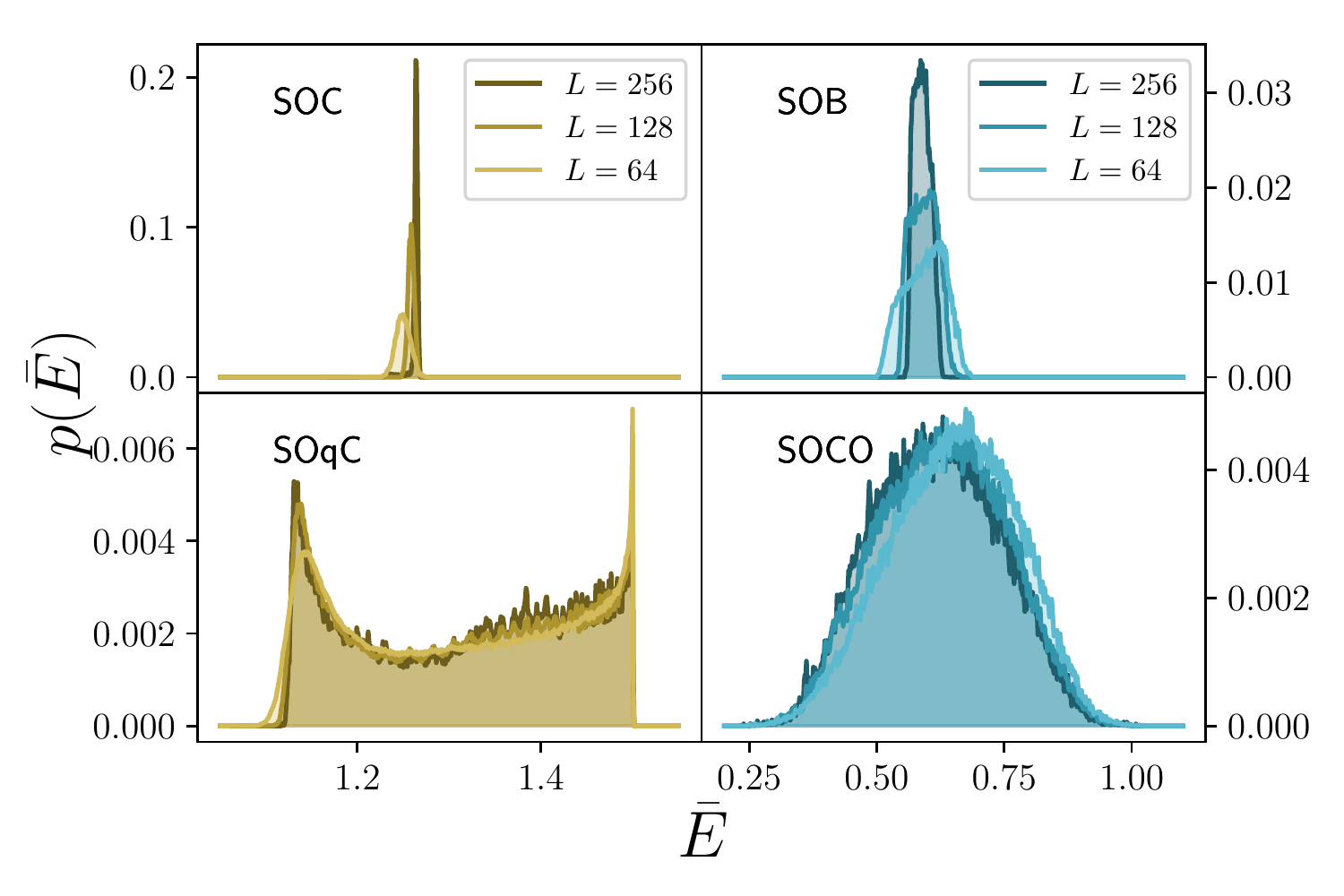}
\caption{\textbf{Distribution of average energy density $\overline{E}$ at the stationary state in the different types of self-organization mechanisms for finite system sizes.} In the case of SOC and SOB (upper panels), the (unimodal) distribution of energy values becomes progressively more peaked at the transition point as the system size is enlarged, converging as $N \rightarrow \infty$ to perfect self-organization to the transition point (either a critical point $E_c$ for SOC, or the Maxwell point $E_M$ for SOB). In the absence of a conservation law, i.e. in the presence of non-vanishing dissipation term (bottom panels), the distribution remains broad even in the $N \rightarrow \infty$ limit, reflecting the presence of excursions of $\overline{E}$ to both sides of the transition: both in SOqC and in SOCO, the system continuously shifts between the active and absorbing regimes, even in the limit of infinitely-large system sizes. In the case of SOqC, the distribution is broad and bimodal. In the case of SOCO, the broad distribution results from the presence of ongoing oscillations from one phase to the other. In all cases, we simulated the corresponding Langevin equations as described in the text (e.g. Eqs. (\ref{FES}) with updating rules (\ref{updateRule})) for SOC, etc.). For the conserved cases, $\epsilon=0$, while for non-conserved ones we employed the ``offline'' charge rules described by Eqs. \ref{updateRule}. See Table \ref{params-edist} for a list of all parameter values.}  \label{edist} \end{center} \end{figure}

As in the mean-field theory, this system of equations converges to Eqs.(\ref{FES}) in the limit $h/\epsilon \rightarrow 0$ (Fig. \ref{Panels}, upper-left panel). Although the equivalence of Eqs.(\ref{FES}) at criticality and its SOC counterpart Eqs.(\ref{SOC}) is very challenging to prove analytically, it has been consistently demonstrated by means of extensive computational analyses \cite{JABO1}.  Such numerical analyses are possible owing to an exact algorithm to integrate this type of Langevin equations with multiplicative (square-root) noise \cite{Dornic} \footnote{Details of the algorithm, a description of an improvement over the original formulation \cite{Shnerb}, and a code for its implementation can be found in Github: \url{https://github.com/pvillamartin/Dornic\_et\_al\_integration\_class}.}.  Figures \ref{Panels}, \ref{edist} and \ref{avalanches} (upper-left panels) show results from the numerical integration of these equations.  In particular, Figure \ref{edist} (upper-left) shows the probability distribution to find the system in a state with average energy density $\overline{E}$ in the SOC version of the dynamics. This distribution becomes progressively more peaked around $E_c$ as the system is enlarged (since dissipation and driving become arbitrarily small as the system size is increased), converging to a Dirac delta function at $\overline{E}=E_c$ in the infinite-system-size limit.

Some aspects of this mapping have generated long-lasting controversies in the past:
\begin{itemize}
\item The first one regards the conclusion of the above theory that the value to which the self-organization mechanism leads the system, $E^{SOC}$, coincides with the critical point of the standard phase transition in the FES model, $E_c$. This was questioned by using a possible counterexample \cite{Fey}. In particular, for the original BTW deterministic sandpile model in some particular types of lattices is was shown that $E_c^{SOC} \neq E_c$ (for example, for a square lattice $E^{SOC}=2.1252...$ \cite{Fey} but the analytical prediction is $E_c= 2.125$ \cite{Caracciolo}, i.e. there is a deviation in the fourth decimal digit). This result, criticized in \cite{comment-Fey,Russians}, stems from the previously-mentioned lack of ergodicity of deterministic sandpiles, and it does not apply to standard stochastic (ergodic) sandpiles, where the equality $E_c^{SOC} = E_c$ has been consistently verified numerically to hold (see e.g. \cite{JABO1}).

\item The second one concerns the universality class of stochastic SOC models. The numerical values of the exponents are close to those of DP, which led some researchers to claim that SOC models (and FES theory) are in the directed percolation class \cite{Mohanty1,Mohanty2,Mohanty3}.  However, the following observations support the existence of a universality class \textit{per se}, the so-called C-DP (\textit{conserved} directed percolation) or Manna class (see e.g. \cite{Dornic}):

\begin{enumerate}
\item  In Eqs.(\ref{FES}) there is an additional equation with respect to  the DP theory that includes a conservation law. The latter constitutes a relevant perturbation in the renormalization group sense at the DP fixed point \cite{FES-PRL}.

\item There is a mapping from SOC to interfaces moving in random media, whose universality is different from DP (as known from numerical as well as from analytical renormalization group approaches; see \cite{Alava-Munoz,JABO-PRL,JABO-cusps,Wiese-mapping} and references therein).

\item Numerical estimates of critical exponent values for this class with one- and two-dimensional systems are distinct from those of DP \cite{Romu-PRL,Pruessner-universality,JABO1,JABO-MZ,Thesis-JA,Ron2014}. Recent large-scale numerical analyses (of the one-dimensional Oslo sandpile \cite{Oslo}) closed the debate even on more firm bases by confirming the discrepancy with the DP scaling and conjecturing rational values for some of the exponents \cite{Grassberger-2016}. As a side note, let us highlight that obtaining critical exponents numerically in SOC is challenging because there is a very slow decay from initial conditions in the background or energy field, which makes observing true asymptotic behavior necessitate large system sizes and long computer simulations \cite{Mohanty3}. Indeed, in the stationary state of SOC and FES, as first pointed out in \cite{Mohanty3} (see also \cite{hyper-Levine,hyper-Ron,Rosalba}) the energy field is ``hyper-uniform'' (i.e. the standard deviation of field values in a region of size $N$ decays faster than $\sqrt{N}$ \cite{hyper0, Grassberger-2016}). Given the critical slow decay of correlations, a convenient strategy to observe numerically clean exponents consists in preparing initial conditions that preserve hyperuniformity (or naturally obtained from the system's dynamics) \cite{Grassberger-2016}.  Another powerful strategy to discriminate between DP and C-DP consists in perturbing the system introducing walls or anisotropy, because systems in the DP class and in the C-DP class respond very differently to these perturbations \cite{JABO-MZ}. Finally, not only critical exponents but also some correlators have been shown to be different in DP and C-DP with remarkable numerical accuracy \cite{ JABO-cusps, Thesis-JA}.  \end{enumerate} \end{itemize}

Another important point is the lingering (and frustrating) lack of a working renormalization group approach to study analytically the large-scale behavior of the C-DP field theory, Eqs.(\ref{FES}).  Thorough attempts to renormalize the theory have been made in the literature (see e.g. \cite{FES-PRL,Frederic,Janssen-Stenull,Pruessner-RG}), but a sound solution to this problem has yet to be found.

Notwithstanding, as already mentioned the C-DP universality class can be exactly mapped into the pinning-depinning transition of linear interfaces moving in a random media \cite{Wiese-mapping}, also called the quenched-Edwards-Wilkinson class \cite{RG-LIM-1}. This mapping enables an additional route to
understanding the scaling features of SOC systems, providing us with an excellent workbench to check for consistency in computational results. Moreover, given that a working (functional) renormalization group solution exists for the interfaces in random media \cite{RG-LIM-1, RG-LIM-2, RG-LIM-3, RG-LIM-4}, this connection could be used as an inspiration for theoreticians to tackle the renormalization problem of Eqs.(\ref{FES}).

In summary, there exists a full stochastic theory of SOC that explains how a mechanism relying on slow driving and dissipation --operating at infinitely separated timescales-- is able to self-organize a system to the edge of a non-equilibrium continuous phase transition. At the critical point of this absorbing state transition, marginal propagation of activity in the form of scale-free outbursts occurs. From a more technical point of view, such a critical point is in the C-DP or Manna class, equivalent to the quenched-Edwards-Wilkinson class, and different from DP.

\section{Theory of self-organized bistability (SOB)} 

SOC describes the self-organization of a system to the edge of a continuous (or second-order) phase transition. Thus, one could wonder whether there exists a similar mechanism for the self-organization of a system to the edge of a discontinuos (or first order) phase transition, with a region of bistability between active and absorbing phases. This idea, recently scrutinized, has led to the concept of ``self-organized bistability (SOB)'' \cite{SOB} (see also \cite{Gil-Sornette}).

Let us start, once again, by considering the minimal form of a discontinuous absorbing-to-active transition in the simplest possible mean-field terms:
 \begin{equation}\dot{\rho}(t) = a \rho - b \rho^2 -\rho^3 
\label{dis-MF} 
\end{equation} where now $b<0$ and $c>0$.  Indeed, as illustrated in Figure \ref{nullclines} (right panel),  the stationary solution of Eq.(\ref{dis-MF}) exhibits a regime of bistability  between an absorbing and an active state. Coupling this dynamical equation to one for an energy field as in SOC, $\dot{E}=h-\epsilon\rho$, introduces a feedback loop that leads the system to exhibit a limit cycle (the loop in Fig.\ref{nullclines}). Indeed, the nullcline of this second equation is $\rho=h/\epsilon$ which, for small $h/\epsilon$, intersects the other nullcline at an unstable point, thus leading to the creation of a limit cycle \cite{SOB,PRR}.  Therefore, a mechanism identical to that of SOC is able to self-organize a mean-field system that exhibits a discontinuous transition to generating periodic bursts of activity.

In order to go beyond this mean-field picture, a simple modification of the theory above leads to the following set of Langevin equations describing self-organization to the edge of a discontinuous transition in spatially extended systems \cite{SOB}:
 \begin{equation} \begin{array}{lll}
  \partial_t \rho(\vec{x},t) = (a + \omega E({\vec{x}},t))\rho - b\rho^{2}-c\rho^{3}
  +D\nabla^2\rho+\sigma  \sqrt{\rho(\vec{x},t)}\eta({\vec{x}},t) \\ 
\partial_{t} E({\vec{x}},t) = \nabla^2\rho({\vec{x}},t)  \label{FES-SOB}
\end{array}
\end{equation}
where all the terms are as in Eqs.(\ref{FES}) except the coefficient of the quadratic term, negative here (i.e.  $b<0$), and the additional cubic term (with coefficient $c>0$), which needs to be added to preserve stability.

Numerical integration of Eq.(\ref{FES-SOB}) can be performed using the same integration scheme as with SOC. The system can be initialized with either low or high homogeneous values of the density, $\rho$, which enables the system to reach different homogeneous steady states (provided that $|b|$ is larger than a certain (tricritical) value \footnote{Let us remark that, for relatively small (in absolute value) $b$, the transition becomes continuous even if the mean-field approximation predicts a discontinuous one. As discussed in \cite{Eluding}, fluctuation effects typically soften the discontinuity, shrink bistability regions, and can even alter the order of the phase transition, leading to noise-induced criticality.}), thus confirming explicitly that the fixed-energy equations above exhibit a full region of bistability with hysteresis on two-dimensional lattices \cite{SOB}. In addition, within the bistable region there exists a \textit{Maxwell point} ($\overline{E}=E_M$ at which both phases are equally stable) that defines the edge of phase coexistence. The latter is computationally verified by considering as initial condition half a system in the active state and the other half in the absorbing state; right at $\overline{E}=E_M$, the flat interface separating these two halves does not move on average (i.e. none of the two phases is more stable than the other).

The mechanism enabling self-organization to the edge of bistability (SOB) is constructed, as in SOC, by adding slow (``offline'') driving and boundary dissipation to the previous equations. In particular, the system is set into an absorbing state and is locally perturbed to trigger avalanches of activity, which are eventually dissipated at the system boundaries. By iterating this process, the system self-organizes to values of $\overline{E}$ close to $E_M$ (converging exactly to $E_M$ in the thermodynamic limit). Alternatively, again as in the SOC case, we can obtain a similar behavior
by considering ``online'' driving and dissipation, i.e. by replacing the second equation in (\ref{FES-SOB}) with:
\begin{equation}\partial_{t} E({\vec{x}},t) = \nabla^2\rho({\vec{x}},t)
  -\epsilon \rho({\vec{x}},t) + h(\vec{x},t)
\label{SOB}
\end{equation}
in the limit $h/\epsilon\rightarrow0$ (see Fig. \ref{Panels}, upper-right panel).

\begin{figure}[h]
\begin{center}
\includegraphics[height=8cm]{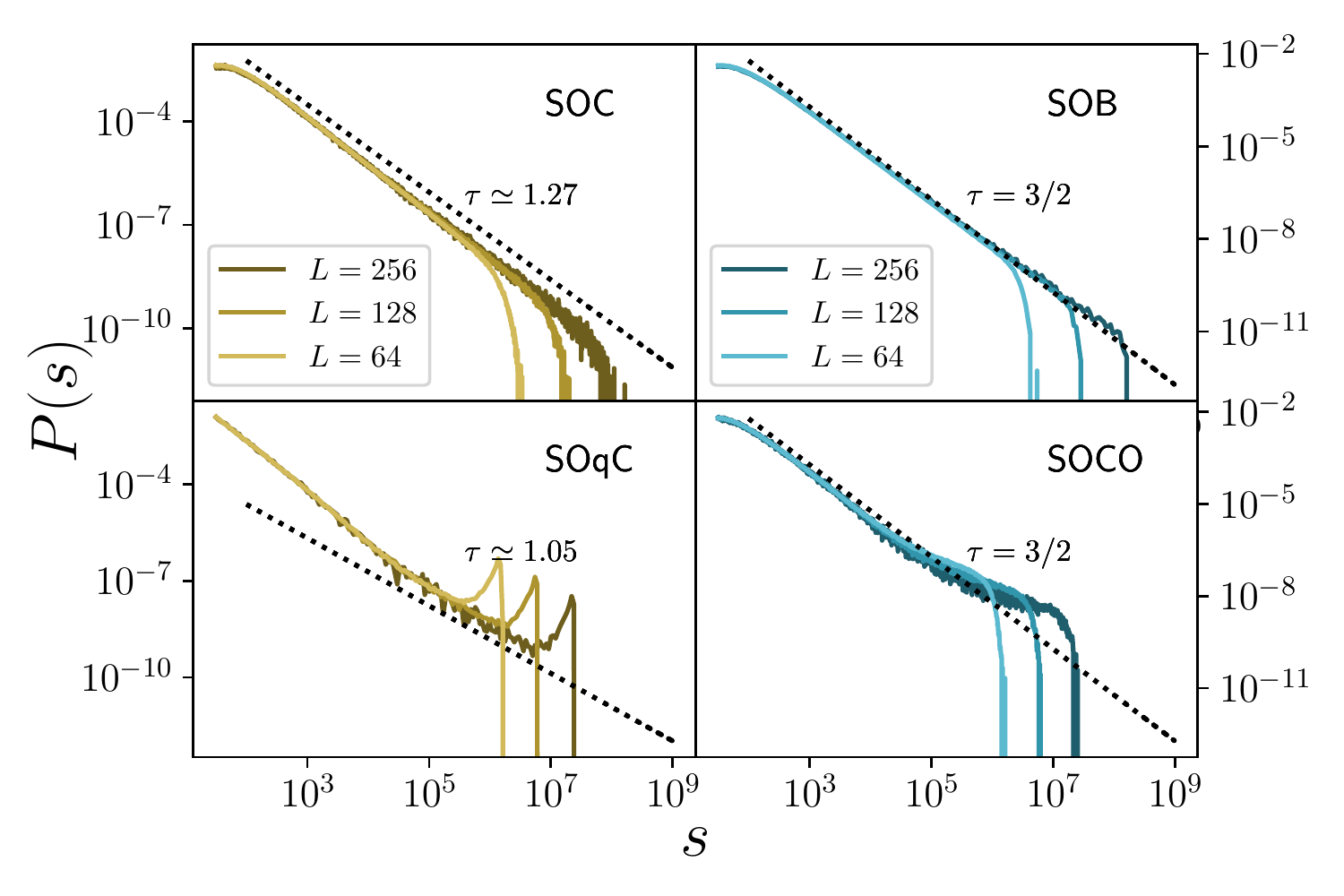}
\caption{\textbf{Avalanche size distributions for the four different types of self-organization mechanisms in two dimensional lattices.} For the cases of SOC and SOB  (upper panels) the distribution can be fitted by a power law in an exact way in the thermodynamic limit, while it is truncated --in a scale-invariant way-- in finite systems. For SOC, the distributions show scaling that belongs to the C-DP or Manna class. In the case of SOB, power-laws are also obtained, and show mean-field exponents (including logarithmic corrections to scaling as the upper critical dimension is $2$, see main text); also, note the bump at the end of the distribution, due to ``king'' events, effect that can be made more apparent by increasing the value of $b$ in the Langevin equation (i.e. making the jump at the discontinuity of the phase transition more abrupt). For the cases of imperfect self-organization (i.e. non-conserved) either SOqC or SOCO (lower panels), the distributions can be fitted by power-laws only in an approximate way. This is a consequence of the fact that, in these cases, the control parameter does not settle to a precise (critical) value but keeps hovering around the edge of the transition even in the thermodynamic limit.
  Parameter values are as in Table \ref{params-edist}, except for SOB, for which $b$ was reduced to $b=-0.7$ to avoid excessively large ``king'' events.}
\label{avalanches}
\end{center}
\end{figure}

Remarkably, avalanches of broadly different scales with signatures of scale invariance also emerge in SOB, in spite of the lack of a critical point \cite{SOB}. However, the avalanche size and duration probability distributions are different from their SOC counterparts in two important ways:

\begin{itemize}
\item The probability distributions for both avalanche size and duration are bimodal: small avalanches coexist with extremely large ones that span the whole system.  These latter ``anomalous'' outbursts of activity, which are also called ``king'' (or ``dragon king'') avalanches in the literature \cite{SOB,Sornette-dragon}, occur in an almost periodic way. They represent waves of activity that propagate almost deterministically (i.e. ballistically) starting from a localized seed, and span through most of the system until they are dissipated at the open boundaries, leaving the system depleted of ``energy''. Let us also emphasize that such system-wide episodes are reminiscent of what happens in the mean-field counterpart, in which activity cyclically ``waxes and wanes'' the system.

\item Smaller standard avalanches have sizes and durations distributed as power laws with exponents $\tau=3/2$ (size, see Fig. \ref{avalanches} upper-right panel) and $\alpha=2$ (duration). These values coincide with those of the mean-field branching process, which is also equivalent to compact directed percolation and the voter model \cite{Marro,Haye,Henkel,Odor,Avalanches}. This type of scaling emerges because the system becomes self-organized to the Maxwell point $E_M$ (see Fig.\ref{edist}, upper-right panel), where the two phases are equally stable (or ``neutral'' \cite{Neutral1,Neutral2}). In this way, clusters of active sites in a non-active environment are equally likely to expand or shrink through fluctuations; this marginality is tantamount to criticality and generates scale invariance.  In this sense, the system behaves as an effective voter model (or compact directed percolation) with two symmetric states in which none of them is favored. Indeed, the voter model exhibits a critical point for the propagation of activity with the
mean-field behavior mentioned above. In two dimensions, upper critical dimension for these systems, logarithmic corrections to scaling appear \cite{Henkel,Dornic-voter}.
 \end{itemize}

 As discussed in detail in \cite{SOB}, the larger the value of $|b|$ --which defines the jump or discontinuity at the phase transition-- the stronger the weight and frequency of anomalous avalanches. Thus, for relatively small jumps, clean scaling can be observed for many decades (as in Fig. \ref{avalanches}, upper-right panel), while for large values of $|b|$ the statistics are more prominently dominated by large anomalous avalanches. In the latter case, larger system sizes are needed to observe clearly the power-law scaling of standard avalanches.
 
 For the sake of completeness, let us mention that it is also possible to construct sandpile models with SOB phenomenology \cite{SOB}. The key difference with respect to standard SOC sandpiles is the presence of a ``facilitation'' mechanism such that activity (i.e. sites above threshold) amplifies in a non-linear way the creation of additional activity. This type of facilitation mechanism is well-known to be at the origin of discontinuous transitions, leading to bistability (see e.g. \cite{Eluding}). The phenomenology of sandpiles with facilitation coincides remarkably well with what we just described for SOB; in particular, they exhibit scale-free avalanches with mean-field exponents together with king avalanches \cite{SOB}. Moreover, in experimental results for real-life sandpiles \cite{Experiment-sandpile} small avalanches coexist with much larger ones, the global energy experiences large excursions, and the empirically determined avalanche distributions are remarkably similar to those of SOB.  Furthermore, it seems that inertia in the dynamics of real sandgrains plays a role similar to facilitation.  All these observations together suggest that SOB is potentially a more adequate theory to describe real sandpiles than SOC. Similarly, SOB could also be at the origin of the ``self-organized avalanche oscillator'' found in microfracture experiments \cite{Papa-Zapperi}.  Finally, in the context of neurodynamics, models of neuronal activity regulated by the level of synaptic resources --very similar in essence to SOB-- can reproduce scale-free avalanches coexisting with anomalous large waves of activity in agreement with empirical observations \cite{Lucilla-dragon} (see next sections for more details on neural dynamics).

\section{Theories for imperfect self-organization}

The theories of SOC and SOB rely heavily onto conservative (bulk) dynamics as well as onto infinite separation of timescales between driving and (boundary) dissipation.  These ingredients, as we have extensively discussed, are essential to achieving a precise and exact self-organization to either a critical point (SOC) or to the point of phase coexistence (SOB).  On the other hand, there is a large variety of natural phenomena that exhibit scale invariance
(at least approximately) 
and in which some form of (bulk) dissipation is inevitably present and/or timescales are not perfectly separated. As an illustrative example, let us discuss the case of neuronal dynamics in the cerebral cortex. Seminal experiments revealed that the dynamics of actual neural networks are bursty, and that critical-like scale-free avalanches of activity can be measured experimentally under generic experimental conditions \cite{BP}. It has been argued that such a critical-like state induces important functional advantages for information processing and transmission in the cortex \cite{Plenz,Chialvo,Gollo,Plenz-functional,Beggs-functional,Healthy,KC,Boedecker} (for a recent review, see \cite{RMP}). This caught the attention of physicists, who readily tried to describe neural networks in terms of SOC \cite{Levina,Lucilla,Millman,Cowan,Gross2014}. 
However, neurons are ``leaky'', as there is no conserved quantity in their dynamics (for instance, the membrane voltage decays spontaneously to some baseline level in the absence of inputs). Moreover, timescales in the brain are not infinitely fast/slow. Therefore, the scaling behavior of cortical networks observed empirically cannot be exactly ascribed neither to SOC \cite{JABO2} nor SOB \cite{PRR}. In order to understand this type of scaling, a more general theory that does not rely on infinite separation of timescales and conservation laws is needed.

Alternative mechanisms for alleged self-organization to criticality in the absence of conservation have long been studied \cite{GG}. Indeed, some of the archetype models of self-organized criticality (other than sandpiles) are non-conserved. Prominent examples are earthquake models \cite{OFC,earthquakes} and forest-fire models \cite{FF1,FF2,FF3}. These are non-trivial models with a rich and complex phenomenology showing power-laws and scaling for at least some decades.  However, the lack of theoretical arguments as solid as the ones discussed above for conserving systems led to a long-standing controversy regarding the existence of true generic scale-invariance in these non-conserving systems. It is not our scope here to review this controversy, but let us just to summarize the main conclusion: none of the studied self-organizing non-conserved models is truly critical but, instead, they exhibit some sort of ``approximate'' or ``relaxed'' criticality (see e.g. \cite{OFC-no1,OFC-no2,Grassberger-FF1,Grassberger-FF2,Grassberger-FF3,Grassberger-FF4,Zierenberg-PRE}, as well as \cite{JABO1} for further discussions and references).

In what follows, we use our unified theoretical framework to briefly introduce and discuss versions of SOC and SOB, respectively, in which the strict conditions of conservation and infinite separation of timescales are relaxed.

\vspace{0.5cm}

\subsection{Theory of Self-organized quasi criticality (SOqC)}
To provide non-conserved systems alleged to be SOC with a general theoretical background, some of us proposed a modified version of the SOC theory, Eqs.(\ref{FES}), that includes explicitly a non-vanishing energy-dissipation term:

\begin{equation} \begin{array}{lll}
  \partial_t \rho(\vec{x},t) = (a + \omega E({\vec{x}},t))    \rho-b\rho^{2}
                   +D\nabla^2\rho+\sigma  \sqrt{\rho(\vec{x},t)} \eta({\vec{x}},t) \\ 
                   \partial_{t} E({\vec{x}},t) = \nabla^2\rho({\vec{x}},t) -\epsilon \rho({\vec{x}},t), \label{SOqC}
                 \end{array}
               \end{equation}
               
that is, Eqs.(\ref{SOC}) with $h=0$, and where now $\epsilon > 0$ is not necessarily small and does not vanish in the large-system-size limit. This equation can be complemented with the following ``offline'' updating rule, inspired in the charging mechanism in models of forest fires and earthquakes \cite{JABO1}:  every time the system reaches the absorbing state, a small ``seed'' of activity is placed at a randomly chosen site, and the energy of all sites is increased:
   \vspace{-0.4cm}
\begin{eqnarray}
  \rho(\vec{x}_0,0) &\rightarrow&  h   \nonumber \\ 
  E({\vec{x}}, 0)   &\rightarrow &E({\vec{x}}, 0) + \gamma (E_\text{max} - \overline{E}) \label{updateRule}
\end{eqnarray} 
where $\gamma$ is an external driving, $E_\text{max}$ the maximum allowed energy in the system, $\overline{E}$ the system average energy density, and $\vec{x}_0$ a random position in the lattice. Note that this ``offline'' updating rule has been used for the $\epsilon\ne0$ cases in Figures \ref{edist} and \ref{avalanches}. These modifications with respect to the SOC case lead to the following results \cite{JABO1}:

\begin{itemize}
\item
First, the leakage (i.e. dissipative) term prevents the existence of a true self-sustained active phase. This can be easily seen by integrating formally the second equation and plugging the result into the first one, thus generating a non-Markovian term $-\epsilon \rho({\vec{x}},t) \int^t_0 dt' \rho({\vec{x}},t')$, which is characteristic of \textit{dynamical percolation} \cite{DyP1,DyP2,DyP3,IAS}. This makes it impossible to have a steady state with $\rho({\vec{x}})\neq 0$ in the long-time limit. Moreover, Eqs.(\ref{SOqC}) exhibit  a transition at some value of the initial energy,  $E_p > E_c$, that separates a \textit{spreading} phase (in which local perturbations of activity can propagate by percolating transiently through the system without reaching a steady state) from a non-spreading phase where  perturbations cannot span the whole system.

\item As a consequence of the previous argument, scaling features in this type of models are related to dynamical percolation when using ``offline'' driving, rather than to C-DP \cite{JABO1}. In other words, bulk dissipation (i.e. breaking the bulk-conservation law) is a relevant perturbation in the renormalization group sense \cite{DyP2}. See Appendix for further details.

\item An analytical and computational study of Eqs.(\ref{SOqC}) revealed that, in this case, increasing $\overline{E}$ through the addition of energy like in sandpiles shifts progressively the systems into the dynamical-percolating phase beyond its critical point $E_p$ \cite{JABO1}. If an avalanche occurs, the associated strong dissipation depletes the system of energy, thus pushing it deep into the non-percolating phase. Therefore, the system does not self-organize exactly to the edge of a phase transition as in the conserved cases above but, instead, it keeps hovering around it, with excursions of finite amplitude to both sides of the (dynamical percolation) transition point, $E_{p}$ (see Figs. \ref{Panels} and \ref{edist}, lower-left panels). In other words, the average energy does not self-tune to a critical value but keeps on
alternating between subcritical and supercritical values, even for infinitely large systems.  Numerical results reveal that this sweeping though the phase transition point might  suffice to induce approximate or ``dirty'' scaling behavior, but not  strict ``\textit{bona fide}'' scale invariance \cite{JABO1,Sornette,Palmieri}. 

\end{itemize}

This mechanism, accounting generically for non-conservative self-organized systems, has been termed ``\textit{self-organized quasi-criticality}'' (SOqC) \cite{JABO1} \footnote{The concept on ``weak criticality'', proposed more recently, bears strong resemblance to SOqC \cite{Palmieri}; see also the slightly different definition of quasi-critical employed in \cite{Rashid}.}. Several remarks are in order:
\begin{itemize}

\item In systems in which driving does not occur ``offline'' (i.e. at an arbitrarily slow timescale, where both the activity and the energy are perturbed only between avalanches) one needs to include explicitly a continuous ``online'' driving term in Eqs.(\ref{SOqC}), so that the second equation becomes $\partial_{t} E({\vec{x}},t) = \nabla^2\rho({\vec{x}},t) -\epsilon \rho({\vec{x}},t)  + h$, where $h$ is the (arbitrarily large) charging or driving rate; alternatively:
\begin{equation}
\partial_{t} E({\vec{x}},t) = \nabla^2\rho({\vec{x}},t) -\epsilon \rho({\vec{x}},t)  + h (E_{max}-E) \label{dissipative2}
\end{equation}
if there is a maximum possible level of charging given by $E_{max}$.  These alternative charging mechanisms may change the previously described phenomenology. The ``online'' driving parameters can be fine-tuned to effectively compensate for dissipation, and a steady state with $\rho\neq 0$ can be achieved. In this case the system phenomenology is controlled by the C-DP transition even if the system does not become truly critical (it just hovers around the critical point, $E_c$); energy is conserved on average, and an approximated or ``dirty'' C-DP-like behavior emerges. However, dynamical percolation dominates for sufficiently large systems if ``offline'' charge is used because, during avalanches, energy can only be dissipated, i.e. bulk conservation is not present during the dynamics. The system will always deplete the available energy until falling again into the absorbing state, when the system is charged to restart the dynamics (see Fig. \ref{avalanches}, lower-left panel).

\item It is important to underline that, in spite of its name reminiscent of SOC, SOqC does not describe true ``self-organization'' to a unique dynamical state. The ratio between dissipation and driving constants $h/\epsilon$ (and also $E_{max}$) determines the system state, thus acting as a true control parameter. If dissipation dominates strongly, the system is subcritical (a case sometimes called ``\textit{self-organized subcriticality}''). If driving is strong, then the system becomes supercritical (``\textit{self-organized supercriticality}'') \cite{JABO1,JABO2}. Finally, for a broad range of intermediate situations, the system hovers around a critical point (``self-organized quasi criticality''). Thus, unlike the SOC case, the choice of parameters (and not only system size) can determine the ``cleanliness'' of the observed scaling behavior.
\end{itemize}

For more detailed explanations of all this phenomenology we refer to \cite{JABO1,JABO2} and, for applications in neuroscience, to \cite{Zhou-oscillations,Copelli1, Copelli2,Kinouchi1,Kinouchi2,Kinouchi-quasi,Priesemann,Zierenberg-PRX,Das-Levina,Alireza,Afshin}.

\vspace{0.5cm}

\subsection{Theory of self-organized collective oscillations (SOCO)}

To close the loop, we now discuss self-organization in the case of non-conservative systems exhibiting a discontinuous phase transition (see Figs. \ref{Panels}, \ref{edist}, and \ref{avalanches}, lower-right panels).

A theory for this case can be written combining the activity equation in Eqs.(\ref{FES-SOB}) with a second equation analogous to Eq.(\ref{dissipative2}) for the non-conserved energy (``online'' driving and dissipation). Alternatively, the ``online'' driving component can be replaced by the (``offline'') rule in Eqs.(\ref{updateRule}) to ``charge'' between avalanches. However, in order to make the presentation more appealing, we will instead discuss the recently introduced Landau-Ginzburg theory for cortical dynamics in the presence of synaptic resources \cite{PNAS-Serena}, which fits perfectly our purposes here. The theory is defined by the following set of equations (considered on e.g. a two-dimensional lattice \cite{PNAS-Serena}):
\begin{equation} \begin{array}{lll}
  \partial_t \rho(\vec{x},t) = (a + \omega E({\vec{x}},t))\rho - b\rho^{2}-c\rho^{3}
  +D\nabla^2\rho+\sigma  \sqrt{\rho(\vec{x},t)}\eta({\vec{x}},t) \\ 
\partial_{t} E({\vec{x}},t) =\nabla^2\rho({\vec{x}},t) - \epsilon E\rho  + h (E_{max}-\overline{E})      \label{PNAS}
\end{array}
\end{equation}
with $b<0$ and $c>0$. In the context of neural dynamics, $\rho(\vec{x},t)$ represents the density of neuronal activity in a coarse-grained region of the cortex, while the energy field represents the level of synaptic resources at a given location (with $E_{max}$ its maximum level at any given location).  These equations are similar to those for SOB, Eqs.(\ref{FES-SOB}), but note the presence of a dissipative term in the second equation (similar to, but different from, that in the SOqC theory, Eq.(\ref{SOqC})), as well as a driving term (as in Eq.(\ref{dissipative2})) that ``charges'' the energy field. The diffusion term in the second equation could be safely removed as it is irrelevant in this case \cite{PRR}, and was actually absent/omitted in the original neural-dynamic model \cite{PNAS-Serena}.

As commented for SOqC, because the dynamics are not conserved, the system is not really ``self-organized'' to a unique type of behavior \cite{PNAS-Serena}. Indeed, the free parameter $E_{max}$ becomes a control parameter, regulating the system output:

\begin{itemize}
\item If $E_{max}$ is exceedingly small, the system ``self-organizes''  into an absorbing configuration with no activity.
 
\item If $E_{max}$ is sufficiently large, the system ``self-organizes'' into a homogeneous active state where individual sites alternate between the active and the inactive state; the latter occurs in an incoherent or ``asynchronous'' way, thus keeping an overall fixed stationary density of activity.

\item In the more interesting case between the two regimes above, there is an intermediate phase in which quasi-oscillatory dynamics emerge. This regime is described by waves of activity traveling through the system, generating co-activation of many units within a relatively small time window (we refer to \cite{PNAS-Serena} for more details and videos of these rich dynamics). These events bear strong resemblance with the system-spanning avalanches --or anomalous waves-- described in SOB.
\end{itemize}

By fine-tuning $E_{max}$, it is possible to find a critical point that separates the phase of global oscillations (``synchronous phase'') from the active phase in which units do not oscillate in unison (``asynchronous phase''). In other words, these systems exhibit  a synchronization phase transition \cite{PNAS-Serena}. 

Finally, let us remark that, in the limit in which the driving and dissipation parameters $\epsilon$ and $h$ converge to $0$ (keeping the usual separation of timescales), the system approaches true self-organization.  Not surprisingly, in this limit one recovers all the phenomenology of SOB, including scale-free avalanches coexisting with anomalously large waves of activity \cite{PRR}.

\section*{Summary and Discussion}

More than three decades after the creation of the concept of self-organized criticality, SOC continues to attract interest of theoretical and applied scientists. The original prototypical models such as sandpiles rely on a rather general type of feedback mechanism that, acting differentially at both sides of the phase transition, allows for the self-organization to the edge of the transition. As profusely discussed here, such a feedback mechanism depends crucially on a large separation of timescales between a slow driving and the intrinsic fast dynamics, conserved in the bulk. Note that the feedback mechanism is ``just'' a way to reach the neighborhood of a phase transition, but it is the intrinsic dynamics that determines the universality class that the system belongs to. Thus, there is no ``self-organized universality class'', but instead phase transitions that belong to specific universality classes (BTW, C-DP...) and that may be  reached through the described self-organization mechanism.

Although other mechanisms for self-organization to criticality that do not depend on such a type of feedback were originally proposed (e.g. extremal dynamics \cite{Bak-Sneppen}), in this mini-review paper we have focused instead on this feedback mechanism to provide the reader with a concise and systematic overview of field theoretical or, equivalently, Langevin approaches to SOC. This formalism --in the spirit of Landau-Ginzburg and Hohenberg-Halperin-- constitutes, in our opinion, an excellent framework to underline the generality of the discussed phenomenology, stressing the key aspects and neglecting as much as possible specific model-dependent details.

Thus, we reviewed the Langevin approach to SOC and described how and why the system self-organizes to the edge of a standard (non-equilibrium) continuous phase transition separating an active from an absorbing phase. In the limit of an infinite separation of timescales and conservative bulk dynamics, the systems self-organizes perfectly to the phase transition, i.e. to criticality. On the other hand, when some of these stringent conditions are relaxed (i.e. if the separation of timescales is not perfect and/or the system is not perfectly conservative), then there is instead approximate or ``imperfect'' self-organization to the vicinity of the transition point, with the system's control parameter hovering around it and excursions into both the subcritical and the supercritical phases (SOqC). Forest-fire and earthquake models --as well as models of neural dynamics-- can much better be ascribed to SOqC than to actual SOC. It is, however, important to underline that tuning the parameters associated with driving and dissipation is required for the system to self-organize either to the subcritical or the supercritical regimes. Thus, SOqC systems do not really self-organize to the vicinity of a transition in a strict sense, but rather there are broad ranges of parameter values for which the system hovers around criticality and exhibits approximate scale invariance.

We also reviewed the recently proposed concept of SOB, explaining how a feedback mechanism similar to that of SOC may operate to self-organize a system to the edge of a first-order, discontinuous, phase transition. As for SOC, in the limit of infinite separation of timescales and conservative bulk dynamics, the self-organization to the transition is exact. Unlike for SOC, however, small avalanches coexist with anomalously large ones. Furthermore, avalanches belong to the voter class universality class, which results from the existence of bistability (i.e. two equivalent states as in the voter model class) at the self-organized Maxwell point. We also defined an ``imperfect'' self-organization mechanism for a family of systems exhibiting a discontinuous phase transition. As in SOqC, there is not ``true'' self-organization. Instead, the non-conserved equivalent of SOB shows a broad range of parameter values for which the system exhibits collective oscillations, alternating between regimes of high activity and quiescent ones (hence the name ``self-organized collective oscillations'', SOCO).
  
All these mechanisms lead to the self-organization to the edge or vicinity of a non-equilibrium phase transition.
Nevertheless, similar mechanisms have also been described in other contexts, such as self-organization to the edge of a synchronization phase transition in the context of models of neuronal dynamics \cite{STDP-Sisyphus,STDP-Mirasso,STDP-Afshin}. This self-organization mechanism is similar in spirit to those above, operating differentially in the two competing phases. In particular, the synaptic strengths (which play the role of ``energy variable'') tend to be reinforced when the system is in the asynchronous phase and weakened when it is exceedingly synchronous (which is achieved by a synaptic plasticity mechanism such as``spike-time dependent plasticity'' \cite{STDP}).  In this way the system can be kept self-organized nearby the edge of synchronization. 

In summary, we have reviewed within a common and unified framework different types of mechanisms  for the self-organization to the the vicinity of phase transitions. We hope that this work help clarify the --sometimes confusing or contradictory-- literature on the subject, and contribute to pave the road for new and exciting developments in physics, but also other disciplines. This could be especially important in biology, where the idea that living systems can obtain important functional advantages by operating at the edge of two alternative/complementary types of phases/states has attracted a great deal of attention and excitement \cite{RMP}. The ideas discussed here can be re-interpreted in terms of homestatic mechanisms allowing living systems to regulate themselves to operate in the desired operational regimes.


\appendix
\renewcommand{\thesubsection}{\Alph{subsection}}
\renewcommand\thefigure{\thesubsection}    
\renewcommand\thetable{\thesubsection}    

\section*{Appendices}

\subsection{Table with parameter values}
\setcounter{table}{0}    

\begin{table}[h]
\begin{center}
\begin{tabular}{| c | c | c | c | c |} 
 
 \hline
 \textbf{Parameter} & \textbf{SOC} & \textbf{SOqC} & \textbf{SOB} & \textbf{SOCO} \\
 \hline   
 $a$       & $-1.00$ & $-1.00$ & $-1.00$ & $-1.00$   \\ 
 \hline
 $b$       & $1.00$  & $1.00$  & $-1.50$ & $-1.50$  \\
 \hline
 $c$       & -  & -  & $1.50$ & $1.50$  \\
 \hline
 $\omega$  & $1.00$ & $1.00$ & $1.00$ & $1.00$    \\
 \hline    
 $h$       & $1.00$  & $0.10$  & $1.00$ & $1.00$  \\
 \hline
 $\gamma$  & $0.00$  & $0.10$  & $0.00$ & $0.02$  \\
 \hline
 $\epsilon$& $0.00$  & $0.10$  & $0.00$ & $0.10$  \\
 \hline
 $E_{max}$ & -  & $1.50$  & - & $1.30$  \\
 \hline
 $D$       & $1.00$  & $1.00$  & $1.00$ & $1.00$  \\
 \hline
 $D_E$     & $1.00$  & $0.10$  & $1.00$ & $1.00$  \\
 \hline
 $\sigma$   & $1.00$  & $1.00$  & $1.00$ & $1.00$  \\
  \hline
\end{tabular}
\caption{\textbf{Parameters used in the numerical simulations for Figs \ref{edist} and \ref{avalanches}.} Dashes indicate that the corresponding parameter is not present in the model.}
\label{params-edist} 
\end{center}
\end{table}

\subsection{C-DP approximate scaling in SOqC}
\setcounter{figure}{0}    

\begin{figure}[h] \begin{center} \includegraphics[height=6cm]{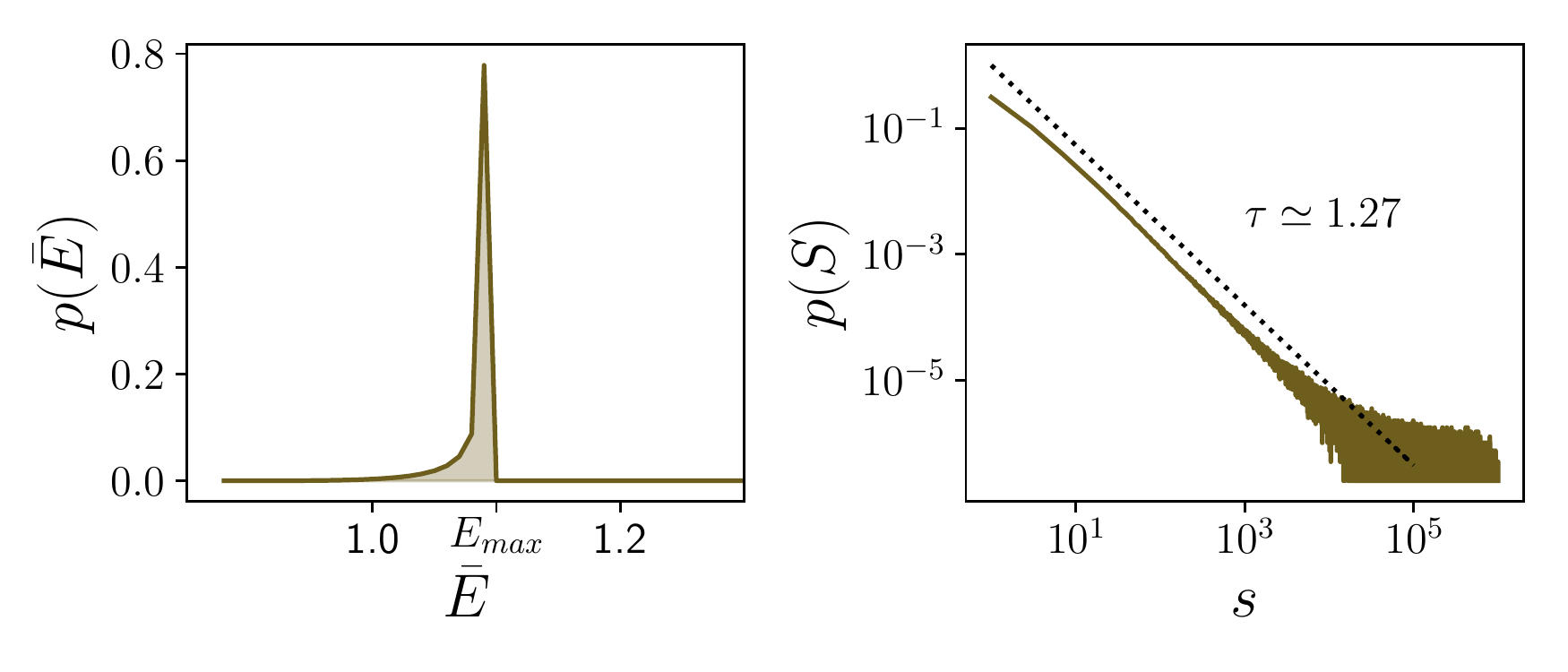} \caption{\textbf{Simulations of the SOqC theory may exhibit scaling similar to that of C-DP.} (A) The probability distribution of the average energy is peaked around $E_c$, but it is much more sprread that the SOC case, owing to oscillations around the critical point. (B) Distributions of avalanche sizes in this case. The C-DP exponent is shown for comparison. Although a change in trend can be seen around $s\sim 10^4$, larger sizes are required in order to clearly see the exponent correspoding asymptotic scaling (controlled by dynamical percolation, as discussed in the main text). Parameter values are: $a=0.423$, $b=\omega=1$, $D=D_E=0.25$, $\sigma=\sqrt{2}$, $\gamma=0.1$, $h=1.0$, $\epsilon=0.1$, $E_{max}=1.1$, $L=256$.}
\label{soqc-cdp}
\end{center}
\end{figure}

Although the case of the SOqC has been argued to be belong, in general, to the dynamical percolation universality class, it is possible to select parameter values such that avalanches present a scaling controlled, at least transiently (i.e. for small sizes and durations) by C-DP  (see Figure \ref{soqc-cdp}, and \cite{JABO1} for more details). For instance, in the case of ``offline'' driving, if  the driving is not strong enough as to bringing the system above the critical point for spreading, the averaged energy $\overline{E}$ hovers around the C-DP critical point $E_c$, as shown in \cite{JABO1}. Actually, there is a value of the charge rate, $\gamma_s$, that allows the system to enter into the spreading phase leading to dynamical-percolation type of scaling. Similarly, ``online'' driving can effectively compensate for dissipation so that an steady state can be reached: as discussed in the main text, this state can be either sub-critical, supercritical, or near critical, depending on the relative strengths of driving and dissipation. In the near-critical case the critical-like features are expected to be
controlled by the C-DP point due to the dynamic ``online'' addition of energy, which perturbs the dynamical-percolation (dissipative) behavior.

\section*{Conflict of Interest Statement}

The authors declare that the research was conducted in the absence of any commercial or financial relationships that could be construed as a potential conflict of interest.

\section*{Author Contributions}

M.A.M., J.A.B., and S.S designed the research. J.A.B. and V.B. performed simulations. V.B. prepared the figures. All authors contributed to writing and reviewing the manuscript.

\section*{Funding}
We acknowledge the Spanish Ministry and Agencia Estatal de
  investigaci{\'o}n (AEI) through grant $FIS2017-84256-P$ (European
  Regional Development Fund(ERDF), as well as the Consejer{\'\i}a de
  Conocimiento, Investigaci{\'o}n y Universidad, Junta de
  Andaluc{\'\i}a and ERDF, Ref.  $A-FQM-175-UGR18$ and $SOMM17/6105/UGR$ and for financial support.
  We also thank Cariparma for their support through the TEACH IN PARMA project.

\section*{Acknowledgments}

We are very thankful to Ronald Dickman, Guillermo B. Morales,  Johannes Zierenberg, Matteo Sireci
for comments and a critical reading of initial versions of the manuscript.

\bibliography{SOC-SOB}

\end{document}